\documentclass[a4paper,12pt]{article}

\usepackage{amsmath}
\usepackage{amssymb}
\usepackage{graphicx}
\usepackage{natbib}
\usepackage{color}
\usepackage{epstopdf} 

\usepackage{setspace} 

\date{}

\begin{document}

\begin{flushleft}
{\Large
\textbf{mRNA diffusion explains protein gradients in  \textit{Drosophila} early development}
} \\
Rui Dil\~ao$^{1,\ast}$, 
Daniele Muraro$^{1}$
\\
\bf{1} Nonlinear Dynamics Group, Instituto Superior T\'ecnico
Av. Rovisco Pais, 1049-001 Lisbon, Portugal
\\
$\ast$ E-mail: Corresponding rui@sd.ist.utl.pt
\end{flushleft}

\section*{Abstract}
We propose a new model describing the production  and the establishment of the stable gradient of the Bicoid protein along the antero-posterior  axis of the embryo of \textit{Drosophila}.  In this model,  we consider that \textit{bicoid} mRNA   diffuses along the antero-posterior axis of the embryo and  the protein is produced in the ribosomes localized near the syncytial nuclei. Bicoid protein stays localized near the syncytial nuclei as observed in experiments.
We calibrate the parameters of the  mathematical model with experimental data taken during the cleavage stages $11$ to $14$ of the developing embryo of \textit{Drosophila}.  We obtain  good  agreement between the experimental and the model gradients, with relative errors in the range $5-8\%$. The inferred diffusion coefficient of  \textit{bicoid} mRNA  is in the range $4.6\times 10^{-12}-1.5\times 10^{-11}$~m$^2$s$^{-1}$, in agreement with the theoretical predictions and experimental measurements for the diffusion of macromolecules in the cytoplasm.  We show that 
the model based on the mRNA diffusion hypothesis is consistent with the known observational data, supporting the recent experimental findings of the gradient of \textit{bicoid} mRNA  in \textit{Drosophila} [Spirov \textit{et al.} (2009) \textit{Development} 136:605-614].

\section*{Introduction}

In \textit{Drosophila} early development,   \textit{bicoid} mRNA  of maternal origin is deposited in one of the poles of the egg, determining the anterior tip of the embryo, \citep{Frigerio, Driever}. The deposition of the mRNA is done during oogenesis and is transported into the oocyte along microtubules, \citep{Saxton}. After fertilization and deposition of the egg, and during the first $14$ nuclear divisions of the developing embryo,   \textit{bicoid} mRNA   of maternal origin is translated into protein in the  ribosomes.  

During the interphases following the 11th nuclear division up to the 14th, the concentration of Bicoid protein distributes non-uniformly along the antero-posterior axis of the syncytial blastoderm.
Bicoid has higher concentration  near the anterior pole of the embryo,  and its local concentration decreases as the distance to the anterior pole increases. This is called the Bicoid protein gradient, \citep{Nuss}.   

As, during oogenesis,   \textit{bicoid} mRNA  is deposited near the anterior pole of the embryo,
it is implicitly assumed that  Bicoid protein is produced in the ribosomes of the nuclei localized near the anterior pole of the embryo, and then diffuses through the syncytial blastoderm.  Eventually, this protein diffusion could be facilitated by the absence of cellular membranes in the syncytial phase of the developing embryo.  Driever and N\"usslein-Volhard \citep{Driever} argued that the protein gradient is generated by protein diffusion and degradation throughout the embryo. Later, N\"usslein-Volhard \citep{Nuss, Nuss2}  emphasized that the Bicoid protein diffuses away from the site of its production,  the local mRNA deposition region. The theoretical possibility of this mechanism of morphogenesis goes back to the work of Alan Turing in the fifties \citep{Turing}, and has been further discussed and analyzed by Wolpert \citep{Wolpert}, Crick \citep{Crick} and Meinhardt \citep{Meinhardt}. Experimental measurements within mammalian cells \citep{Wojcieszyn, Mastro} and theoretical analysis \citep{Crick} suggested that diffusion coefficients of macromolecules in the cytoplasm are in the range $10^{-11}-10^{-13}$~m$^2$s$^{-1}$.

\begin{figure}[!ht]
\begin{center}
\includegraphics[width=4.7in]{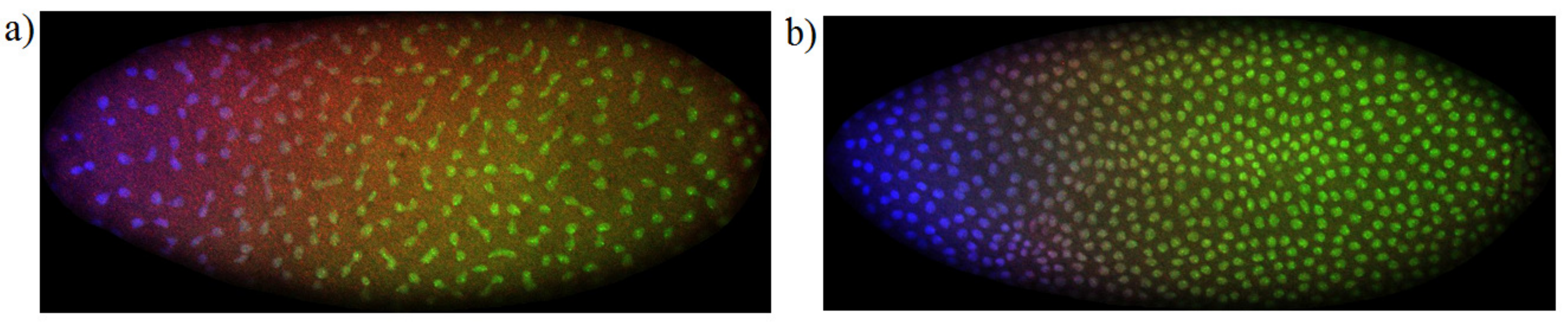}
\end{center}
\caption{
{\bf Distribution of Bicoid protein (in blue) in the embryo of \textit{Drosophila}, in the interphase following the cleavage stages 11 (a) and 12 (b).}  The images are  from the FlyEx datasets ab18 (a) and ab17 (b), \citep{FlyEx1,FlyEx2,FlyEx3, Poustelnikova,Pisarev}. Note the absence of Bicoid protein in the inter-nuclear regions of the cytoplasm.  The localization of Bicoid protein near the nuclear envelopes suggest that ribosomes are also localized near the nucleus.
}
\label{fig1}
\end{figure}

However,
there are several open questions related with the establishment of the stable gradient of protein Bicoid along the antero-posterior axis of the embryo of \textit{Drosophila}. Experimental observations during cleavage stages $11-14$, and before cellularization that occurs at the end of cleavage stage $14$, show that Bicoid protein is always localized around the syncytial nuclei, Figure~\ref{fig1}. This can be seen in the embryo data sets b18, ab17, ab16, ab12, ab14, ab9, ad13, ab8 of the FlyEx database 
\citep[http://flyex.ams.sunysb.edu/flyex/]{FlyEx1,FlyEx2,FlyEx3,Poustelnikova,Pisarev}. This fact suggests that dispersal effects driven by molecular collisions with Bicoid protein --- Brownian motion --- do not play a significant role in the establishment of the gradient of Bicoid.  
As Brownian motion is the driven mechanism of 
diffusion dispersal, \citep[chap. 9]{Murray},
it is difficult to understand how the diffusion of a protein produces a strongly localized protein concentration  around the syncytial nuclei and during successive mitotic cycles. On the other hand, as Bicoid protein is produced and attains a steady state during the first cleavage cycles, its localization near the nuclear envelops suggest that ribosomes are also localized near the nucleus. If ribosomes were not localized near the nuclear envelopes of the syncytial nucleus, protein in the inter-nuclear regions of the embryo would be observed.

As argued by Kerszberg and Wolpert \citep{Kerszberg}, there is not a clear experimental evidence of protein degradation, a necessary mechanism for the establishment of a steady protein gradient in models based on protein diffusion. 

Houchmandzadeh \textit{et al.} \citep{Houchmandzadeh2}, reported the constant Bicoid protein concentration during cleavage cycles 12-14, suggesting the stability of protein concentration during an important developmental period.
On the other hand, 
the protein diffusion hypothesis lead to some quantitative
contradictory facts. For example, in
recent experiments, the hypothetical inferred cytoplasmic diffusion coefficient of the  Bicoid protein during the cleavage stage $13$ is of the order of $0.3\times 10^{-12}$~m$^2$s$^{-1}$, \citep{Gregor}. However, 
during the first cleavage stages of the developing embryo,
the  Bicoid  protein  reaches a steady state in $90$~minutes (end of cleavage stage 9), \citep{Gregor}, and a simple estimate with the Houchmandzadeh \textit{et al.} \citep{Houchmandzadeh} model shows that the diffusion coefficients must be of the order of $2\times 10^{-12}$~m$^2$s$^{-1}$, \citep{Gregor}. This value for the diffusion coefficient is one order of magnitude larger than the value inferred from experiments. This discrepancy  between model estimates and observation needs a clear explanation, \citep{Reinitz}.

Here, with a mathematical model, we show that the observed gradient of the Bicoid protein can be explained by the diffusion of \textit{bicoid} mRNA, and Bicoid protein stays localized near the nuclei of the syncytial blastoderm of the embryo of \textit{Drosophila}. This explains the absence or the very low level of Bicoid concentration in the regions between the nuclei during the first stage of development of \textit{Drosophila}. We determine a scaling relation between the mRNA diffusion coefficient, the embryo length and the mRNA degradation rate, enabling the precise determination of the diffusion coefficient of \textit{bicoid} mRNA.
In this model, it is not necessary to introduce the morphogen degradation hypothesis for protein, and the steady gradient of protein is reached after the complete translation of mRNA of maternal origin.

The  mRNA localization mechanism in the embryo of \textit{Drosophila} has been analyzed  experimentally by several authors, and \citep{Saxton} argues that the relative small size of mRNA suggests that random diffusion and specific anchoring to the cytoskeleton in a target area might suffice
for localization in the syncytial blastoderm. Cha \textit{et al.} \citep{Cha} reported rapid saltatory movements in injected  \textit{bicoid}  mRNA in the embryo, followed by dispersion without localization. Other effects of diffusing mRNA has been reported by Forrest and Gavis \citep{Forrest} 
for the  \textit{nanos}  mRNA. More recently, Spirov \textit{et al.} \citep{Spirov} have shown that a \textit{bicoid} mRNA gradient exists along the antero-posterior axis of the embryo of \textit{Drosophila}, completely changing our current views of this \textit{Drosophila} developmental pathway. The model presented here corroborates these experimental facts, is consistent with the experimental facts and observations, and fits the experimental data with high accuracy.

\section*{Results}

We now derive a mRNA diffusion model and we show that experimental protein gradients are well fitted in this framework. This shows that a mechanism of mRNA mobility (diffusion) is enough to explain protein gradients.

\subsection*{A mRNA diffusion model}
It is an experimental fact that  mRNA of maternal origin is deposited in a small region of the embryo of \textit{Drosophila}, defining the anterior pole of the fertilized egg, \citep{Driever}.
After the deposition of \textit{bicoid}  mRNA, we assume that   \textit{bicoid}  mRNA of maternal origin disperses within the embryo and this process is simultaneous with the successive cleavage stages. Then, the protein is produced in the ribosomes that are  near the nuclear membranes of the nuclei in the syncytium.  Representing 
by $R(x,t)$ the concentration of  \textit{bicoid}  mRNA along the one-dimensional antero-posterior axis ($x$) of the embryo, and by $B(x,t)$  the  concentration of Bicoid  protein, the equations describing the production of Bicoid from mRNA are,
\begin{equation}
\left\{
\begin{array} {ll}\displaystyle
\frac{\partial R}{\partial t}&\displaystyle
=-dR+D \frac{\partial^2 R}{\partial x^2} \\[10pt] \displaystyle
\frac{\partial B}{\partial t}&=aR
\end{array}\right.
\label{eq2.1}
\end{equation}
where $a$ is the rate of production  of Bicoid from mRNA, $d$ is the degradation rate of \textit{bicoid} mRNA, and $D$ is the diffusion coefficient of  \textit{bicoid} mRNA in the cytoplasm. Using the mass action law, this simple model is straightforwardly derived from  the rate mechanisms,
\begin{equation}
\begin{array} {ll}\displaystyle
R\xrightarrow{a}B+R\\ [2pt]
R\xrightarrow{d} 
\end{array}
\label{eq2.1a}
\end{equation}
and then the diffusion term is added to the mRNA rate equation. In general, 
one molecule of mRNA can produce more than one molecule of protein, implying that 
$d<a$. If one molecule of mRNA produces one molecule of protein then, in the mean, we have $d=a$. 

We consider that the length of  the antero-posterior axis of the embryo is $L$, and so $x\in[0,L]$. We take zero flux  boundary conditions, $\frac{\partial R}{\partial x}(x=0,t)=\frac{\partial R}{\partial x}(x=L,t)=0$, and $\frac{\partial B}{\partial x}(x=0,t)=\frac{\partial B}{\partial x}(x=L,t)=0$, for every $t\ge 0$. The protein initial condition is $B(x,t=0)=0$, and   the initial distribution of mRNA is,  
\begin{equation}
R(x,t=0)=\left\{
\begin{array} {lcl}\displaystyle
A>0 & \mbox{if} & 0\le \ell_1\le x\le \ell_2\le L \\[10pt] \displaystyle
0 &   & \mbox{otherwise}
\end{array}\right.
\label{eq2.2}
\end{equation}
where $A$, $\ell_1$ and $\ell_2$ are constants. The function $R(x,t=0)$ describes the initial distribution of \textit{bicoid} mRNA of maternal origin deposited in the region of the embryo $[\ell_1,\ell_2]\subset [0,L]$. The concentration of mRNA of maternal origin deposited in the embryo is then $A(\ell_2-\ell_1)$.
In this model, \textit{bicoid} mRNA  has a fixed initial concentration, and the Bicoid  protein does not degrade.

Equation (\ref{eq2.1}) with the initial condition (\ref{eq2.2}), and the zero flux boundary conditions define the mRNA diffusion model. This model is linear, and has  solutions that can be determined explicitly. Now,  we will show that, within this simple model,  Bicoid  protein attains a gradient like steady state along the embryo.

By standard Fourier analysis techniques, see for example \citep{Alves} or \citep{Haberman}, 
the solution of the first equation in (\ref{eq2.1}) is,
\begin{equation}
\begin{array}{ll}
R(x,t)&\displaystyle =A\frac{\ell_2-\ell_1}{L}e^{-dt}\\[6pt]
&+\displaystyle 2A\sum_{i=1}^{\infty}
\frac{e^{-dt-\frac{n^2\pi^2}{L^2}Dt}}{n\pi}
 \cos\left(\frac{n\pi x}{L}\right)
\left( \sin\left(\frac{n \pi \ell_2}{L}\right)-\sin\left(\frac{n\pi \ell_1}{L}\right) \right)
\end{array}
\label{eq2.3}
\end{equation}
The solution of the second equation in (\ref{eq2.1}) is,
\begin{equation} 
B(x,t)=B(x,t=0)+a\int_0^tR(x,s)\, ds
\label{eq2.4}
\end{equation}
In the limit $t\to \infty$, the equilibrium or steady solution of the Bicoid protein is  calculated from (\ref{eq2.3}) and (\ref{eq2.4}), and we obtain,
\begin{equation}
\begin{array}{ll}  \displaystyle
B_{eq}(x)& \displaystyle=a_1\frac{\ell_2-\ell_1}{L}\\ [6pt]
&\displaystyle +
 2a_1 \sum_{i=1}^{\infty}
\frac{1}{n\pi+\frac{n^3 \pi^3}{a_2^2} }\cos\left(\frac{n\pi x}{L}\right)
\left( \sin\left(\frac{n \pi \ell_2 }{L}\right)-\sin\left(\frac{n\pi\ell_1}{L}\right) \right)
\end{array}
\label{eq2.5}
\end{equation}
where,
\begin{equation}\displaystyle
a_1=A\frac{a}{d}\, ,\  a_2^2=d\frac{L^2}{D}
\label{eq2.5a}
\end{equation}
and we have introduced into (\ref{eq2.4}) the protein initial condition $B(x,t=0)=0$.
A simple calculation shows that the solution (\ref{eq2.5}) can be written as, \citep{Alves},
\begin{equation}
\begin{array}{ll}  
B_{eq}(x)&=\displaystyle 2\frac{a_1}{e^{2a_2/L}-1}\cosh(a_2 \frac{x}{L})\left(\sinh(a_2 \frac{\ell_2}{L})-\sinh(a_2 \frac{\ell_1}{L}) \right)\\[6pt]
&\displaystyle+\frac{a_1}{2}\left(e^{-a_2(x+\ell_1)/L}-e^{-a_2(x+\ell_2)/L} \right)+I(x)
\end{array}
\label{eq2.6}
\end{equation}
where,
\begin{equation}
I(x)=\left\{ \begin{array}{ll} 
a_1\left(e^{-a_2(\ell_1-x)/L}-e^{-a_2(\ell_2-x)/L} \right)/2, & \textrm{if}\ \ x<\ell_1 \\
a_1-\frac{a_1}{2}\left(e^{-a_2(x-\ell_1)/L}+e^{-a_2(\ell_2-x)/L} \right), & \textrm{if}\ \ \ell_1\le x\le \ell_1 \\
a_1\left(e^{-a_2(x-\ell_2)/L}-e^{-a_2(x-\ell_1)/L} \right)/2, & \textrm{if}\ \ x>\ell_2  
\end{array} \right.
\label{eq2.7}
\end{equation}

The steady solution (\ref{eq2.6})-(\ref{eq2.7}) describes the gradient of the Bicoid protein along the antero-posterior axis of the embryo of \textit{Drosophila}. This solution depends on the set of five parameters $a_1$, $a_2$, $L$, $\ell_1$ and $\ell_2$, to be calibrated with experimental data.   The constants $a_1$ and $a_2^2$ given by (\ref{eq2.5a}) define scaling relations of the embryo.

In order to compare the model predictions with the experimental data, the next step is to calibrate the parameters of the mRNA diffusion model (\ref{eq2.6})-(\ref{eq2.7}) with the available experimental data for the gradient of the Bicoid protein.

\subsection*{Calibration of the mRNA diffusion model with the experimental data}
To calibrate the parameters of the mRNA diffusion model (\ref{eq2.6})-(\ref{eq2.7}) with the experimental data, we   use  the data available in the FlyEx database, \citep{Poustelnikova, Pisarev}. We considered Bicoid gradients for cleavage stages 11-14, and we make the additional assumption that during these cleavage stages, the concentration of the Bicoid  protein is in the steady state, \citep{Houchmandzadeh2}.

\begin{figure}[!ht]
\begin{center}
\includegraphics[width=11cm]{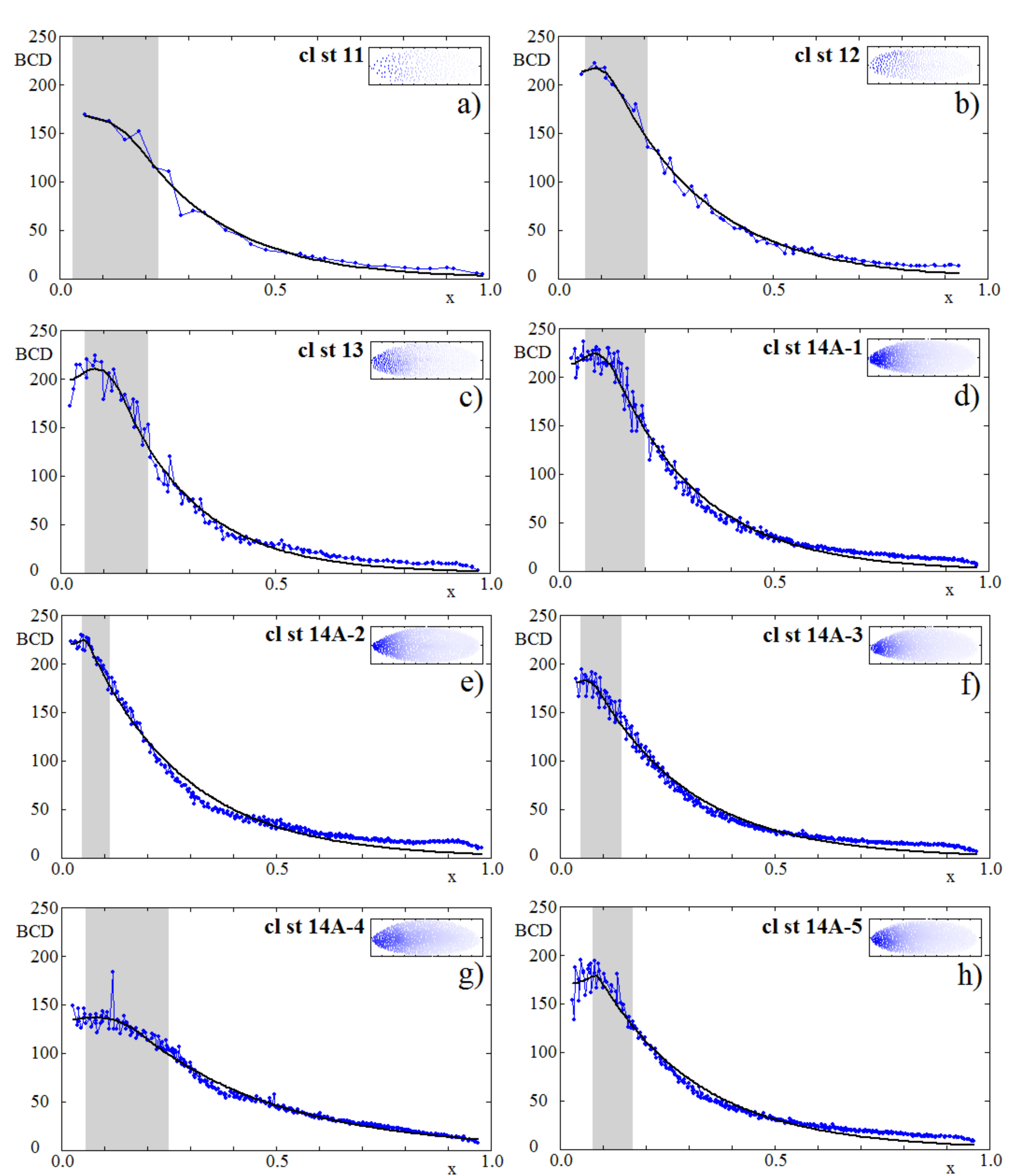}
\end{center}
\caption{
{\bf Concentration of  Bicoid protein (BCD) along the antero-posterior axis ($x$) of \textit{Drosophila}, in the interphase following the consecutive cleavage stages (cl st) 11, 12, 13 and 14.} Data (connected dots) in the figures are from the FlyEx datasets: 
a) ab18 (11); b) ab17 (12); c) ab16 (13); d) ab12 (14A-1); e) ab14 (14A-2); f) ab9 (14A-3); g) ad13 (14A-4); h) ab8 (14A-5), and the numbers inside the parenthesis refer to the cleavage stage, \citep{FlyEx1,FlyEx2,FlyEx3,Poustelnikova,Pisarev}.
In the upper right corner of the figures, we show the corresponding gradients of  Bicoid protein in the two-dimensional projections of the embryo. The data points of the Bicoid gradients are taken from a  region centered around the central antero-posterior axis of the embryo. The transversal length of this region is equal to
$10\%$ of  the maximal length of the dorso-ventral direction.
The lengths of the embryos have been rescaled to the value $L=1$.
The thick black lines are the best fits of the experimental data with the theoretical prediction (\ref{eq2.6})-(\ref{eq2.7}). The gray regions show the initial localization of \textit{bicoid} mRNA, and are defined by the fitted values $\ell_1/L$ and $\ell_2/L$. The parameters of the fits are shown  in Table~\ref{tab:t1}.
}
\label{fig2}
\end{figure}

We have fitted the data sets
of the FlyEx database with the equilibrium distribution of  Bicoid  protein given by  (\ref{eq2.6})-(\ref{eq2.7}).
The fitted functions are represented in Figure~\ref{fig2}.  
In this figure, we show  
the concentration of  Bicoid  protein  along the antero-posterior axis of \textit{Drosophila}, for several embryos and in consecutive developmental
stages. 
Due to the particular form of model prediction (\ref{eq2.6})-(\ref{eq2.7}), the fitted parameter values are $\vec \alpha=(a_1,a_2,\ell_1/L, \ell_2/L)$. The parameter values of the different data sets are shown in Table~\ref{tab:t1}, and correspond to the global minima of the fitness function $\chi^2(\vec \alpha)$, introduced below in (\ref{eq3.1}). 

\begin{table}[!ht]
\caption{
\bf{Fitted model parameters for the protein Bicoid antero-posterior distributions}}
\begin{tabular}{ll|cccc|cc|cc } 
\hline
 &  & $a_1$ & $a_2$ & $\ell_1/L$ & $\ell_2/L$ & $\sqrt{\chi^2_m/B_{max}^2}$ & n& $d$ (s$^{-1}$)   & $Aa(\ell_2-\ell_1)/L$\\ 
\hline
a)&ab18 (11) & 345.2 & 4.69 & 0.03& 0.20 & 0.06 &30&$8.8\times 10^{-4}$ & $5.2\times 10^{-2}$\\ 
b)&ab17 (12)& 894.4 & 4.50 & 0.06& 0.14 & 0.06 &70&$8.1\times 10^{-4}$ & $5.8\times 10^{-2}$\\ 
c)&ab16 (13)& 684.2 & 5.51 & 0.06& 0.15 & 0.08 &152&$1.2\times 10^{-3}$ & $7.4\times 10^{-2}$\\ 
d)&ab12 (14-1)& 927.6 & 4.82 & 0.06& 0.14 & 0.08 &309&$9.2\times 10^{-4}$ & $6.8\times 10^{-2}$\\ 
e)&ab14  (14-2)& 3414.7& 4.38 & 0.05 & 0.07 & 0.08&314&$7.7\times 10^{-4}$ & $5.3\times 10^{-2}$ \\
f)&ab9 (14-3)& 1191.6 & 4.39 & 0.05& 0.10 & 0.07 &343&$7.7\times 10^{-4}$ & $4.6\times 10^{-2}$\\ 
g)&ad13 (14-4)& 470.4 & 3.02 & 0.06 & 0.19 & 0.05 &324&$3.6\times 10^{-4}$ & $2.2\times 10^{-2}$ \\
h)&ab8 (14-5)& 3271.7& 4.25 & 0.08 & 0.09 & 0.07 &332&$7.2\times 10^{-4}$ & $2.4\times 10^{-2}$ \\
\hline 
\end{tabular} 
\begin{flushleft}Parameters values that best fit the experimental distribution of  Bicoid protein shown in Figure~\ref{fig2} with the equilibrium distribution (\ref{eq2.6})-(\ref{eq2.7}). In the first column, we show the data sets and the corresponding cleavage stages.  The  parameters $a_1$, $a_2$, $\ell_1/L$ and $\ell_2/L$ have been determined with the swarm algorithm described in the Materials and Methods section. The lengths of the embryos have been rescaled to the value $L=1$.
$\sqrt{\chi^2_m/B_{max}^2}$ is an estimate of the relative error of the fits, and $n$ is the number of data points in the corresponding graphs in Figure~\ref{fig2}.
The  parameter $d$  has been  determined by (\ref{eq2.5a}) with the estimated diffusion coefficient $D= 10^{-11}$~m$^2$s$^{-1}$ of \textit{bicoid} mRNA and $L=0.5\times 10^{-3}$~m. The parameter $Aa(\ell_2-\ell_1)/L$ has been determined with (\ref{eq3.3a}).
\end{flushleft}
\label{tab:t1}
 \end{table}

The quality of the fits of Figure~\ref{fig2} has been evaluated with the fitness function (\ref{eq3.1}). Denoting by $B_{max}$ the maximum value of each  experimental data set, the mean relative error of a fit is estimated by the quantity, 
$\sqrt{\chi^2_m/B_{max}^2}$, where $\chi^2_m=\min_{\vec \alpha \in S}\chi^2(\vec \alpha)$.
In  Table~\ref{tab:t1}, we show the 
mean relative errors of the fits, and the number of  points ($n$) in each data set.  The  mean relative errors between the theoretical predictions  (\ref{eq2.6})-(\ref{eq2.7}) and the experimental data sets of Figure~\ref{fig2} are in the range $5\%-8\%$, showing a remarkable agreement between the model prediction and the  experimental data. 

To determine the values of the parameter $d$ in Table~\ref{tab:t1},
we have fixed  the embryo length to the  value $L=0.5\times 10^{-3}$~m, \citep[cap. iv]{Nuss2}.  For the 
diffusion coefficient of \textit{bicoid}  mRNA, we have chosen the value, $D=10^{-11}$~m$^2$s$^{-1}$, as estimated below in (\ref{eq3.6b}). By (\ref{eq2.5a}), $d=a_2^2 D/L^2$, and the value of the degradation rate $d$ depends on the choices made for $D$ and $L$. 

As the experimental data is given in arbitrary light intensity units, the initial value of the \textit{bicoid} mRNA concentration is also arbitrary. However. it is plausible to assume that the total amount of initial \textit{bicoid} mRNA deposited in the embryo does not change too much for different embryos. So, in order to estimate the total amount of
\textit{bicoid} mRNA in the embryos,  using the first relation in (\ref{eq2.5a}),
we have calculated the quantity,
\begin{equation} 
Aa(\ell_2-\ell_1)/L=a_1d(\ell_2-\ell_1)/L, 
\label{eq3.3a}
\end{equation}
where $A(\ell_2-\ell_1)$ is the total amount of initial \textit{bicoid} mRNA deposited in the embryo, and the rate $a$ should not change too much for different embryos. Therefore, if the quantity in (\ref{eq3.3a}), does not change too much for different data sets, it is an indication of the ability of the model to describe data sets with different phenotypes. In fact, as shown in Table~\ref{tab:t1}, the quantity (\ref{eq3.3a}) is almost constant among embryos, even if $(\ell_2-\ell_1)/L$ shows a large variability, as is the case of the  fits in Figure~\ref{fig2}.

From the fitted values shown in Table~\ref{tab:t1}, each of the calculated 
parameter values $Aa(\ell_2-\ell_1)/L$, $a_2$ and $d$ have the same order of
magnitude for the different cleavage stages. In fact, the parameters defined in the kinetic mechanisms (\ref{eq2.1a}) are independent of the cleavage stage and, therefore, their values must depend only of the phenotypic characteristics of the analyzed embryo.  The similarities between the parameters $Aa(\ell_2-\ell_1)/L$, $a_2$ and $d$ for different cleavage stages is an indication of the consistency of the theoretical model proposed here.

\subsection*{Determination of the diffusion coefficient of \textit{bicoid} mRNA}

To determine the value of the diffusion coefficient of the \textit{bicoid} mRNA, we use the information that Bicoid protein reaches a steady state in approximately $T$ seconds.   So,  we integrate the two equations in (\ref{eq2.1}) along the embryo length, and using the zero flux boundary conditions, we obtain,
\begin{equation}
\left\{
\begin{array} {ll}\displaystyle
\frac{d {\bar R}}{d t}&\displaystyle
=-d{\bar R} \\[10pt] \displaystyle
\frac{d {\bar B}}{d t}&=a{\bar R}
\end{array}\right.
\label{eq3.4}
\end{equation}
where ${\bar R}$ and ${\bar B}$ are the total amount of mRNA and protein in the embryo, respectively. The  differential equations (\ref{eq3.4}) have the solutions, 
\begin{equation}
\left\{
\begin{array} {ll}\displaystyle
{\bar R}(t)&\displaystyle
={\bar R}(0) e^{-dt}\\[10pt] \displaystyle
{\bar B}(t)&\displaystyle =\frac{a}{d}{\bar R}(0)\left(1-e^{-dt}\right)
\end{array}\right.
\label{eq3.5}
\end{equation}
Assuming  that the steady state of the Bicoid protein is attained after $T$ seconds of development, and that  $95\%$ of the mRNA has been translated into protein, by (\ref{eq3.5}), we have the development time relation, ${\bar B}(T)/(a{\bar R}(0)/d)=0.95=\left(1-e^{-dT}\right)$. From the previous relation we obtain, 
$d=-\log(0.05)/T$. Therefore, by (\ref{eq2.5a}), the diffusion coefficient is,
\begin{equation}
D=d \frac{L^2}{a_2^2}=-\frac{\log(0.05)}{T}\frac{L^2}{a_2^2}
\label{eq3.6}
\end{equation}
As Bicoid protein attains the steady state at the end of cleavage stage 9, in  approximately $T\simeq  90\times 60 $~seconds, \citep{Gregor}, with the data in Table~\ref{tab:t1}, we have $a_2\in [3,5.5]$, and with the choice $L=0.5\times 10^{-3}$~m, by (\ref{eq3.6}), the diffusion coefficient of \textit{bicoid} mRNA is in the range,
\begin{equation}
D\in [4.6\times 10^{-12},1.5\times 10^{-11}]
\label{eq3.6b}
\end{equation}

These estimates, as well as  the numerical fits of Figure~\ref{fig2}, are consistent with the theoretical predictions for the order of magnitude of the diffusion coefficients of large molecules (\textit{bicoid} mRNA) in the cytoplasm \citep{Wojcieszyn, Mastro}.

\section*{Discussion}

We have proposed a new model describing the production  and the establishment of the stable gradient of the Bicoid protein along the antero-posterior  axis of the embryo of \textit{Drosophila}.  In this model, \textit{bicoid} mRNA  diffuses along the antero-posterior axis of the embryo and  Bicoid protein is produced and stays localized near the syncytial nuclei as observed in experiments.

We have calculated the steady state of the  Bicoid  protein along the antero-posterior axis of the embryo of \textit{Drosophila}, 
and we have calibrated the parameters 
of the mRNA  diffusion model with experimental data taken during cleavage stages 11-14.
After the calibration of the model with experimental data, we have predicted the initial localization in the embryo of the \textit{bicoid} mRNA of maternal origin (parameters $\ell_1$ and $\ell_2$ in Table 1), the Bicoid protein concentration profiles along the embryo, and the \textit{bicoid} mRNA degradation rates. The mean relative errors between the theoretical prediction  of the Bicoid protein steady state and the experimental data (Figure~\ref{fig2}) are in the range $5\%-8\%$, suggesting the effective validation of the model proposed here.

A simple estimate gives a   diffusion coefficients $D$ for \textit{bicoid} mRNA  in the interval $[4.6\times 10^{-12},1.5\times 10^{-11}]$~m$^2$s$^{-1}$. This estimate is calculated with the parameters found in the calibration
of the experimental data, with the model prediction formulas for the steady state, and with the additional assumption that the gradient of Bicoid protein is reached at the end of cleavage stage 9, \citep{Gregor}. 
The determination of the diffusion coefficient is strongly dependent of the time duration of the cleavage cycles and therefore, it has a large error that is difficult to quantify.  
On the other hand, as the steady state solution of this \textit{bicoid} mRNA diffusion model depends on the scaling parameter $a_2=\sqrt{dL^2/D}$ and  $a_2$ is determined by
fitting the steady states of the protein profiles, the experimental determination of the degradation rate of  \textit{bicoid} mRNA leads to a more precise estimate of the diffusion coefficient.

The calibration and validation of the mRNA diffusion model shows  that the mechanism of establishment of the gradient of  Bicoid protein observed in \textit{Drosophila} early development can be justified by a diffusion hypothesis for mRNAs.  
The mathematical model considers that \textit{bicoid} mRNA  diffuses along the embryo  and the translated protein stays localized near the syncytial nuclei, as observed in Figure~\ref{fig2}. In this model, protein degradation is not considered and proteins do not diffuse along the embryo. The model proposed here explains the experimental data for protein gradients, and the common 
assumption that morphogen gradients are obtained with a balanced and continuous production and degradation of proteins is not necessary. The low level of Bicoid concentration in the intranuclear regions of the embryo is easily explained through the ribosome localization near the syncytial nuclei.

Random motion of  \textit{bicoid} mRNA  has been observed, \citep{Cha, Saxton}, and the \textit{bicoid}  mRNA gradient has been recently found by Spirov \textit{et al.} \citep{Spirov}. The mechanism of mRNA
diffusion proposed here is uni-dimensional and together with the very good agreement between the experimental data and the model predictions, we can raise the hypothesis of the existence of a mechanism 
of constrained mRNA diffusion along a network of nonpolar microtubules. This
hypothesis has been discussed by Spirov \textit{et al.} \citep{Spirov} and is consistent with a mechanism based on mRNA diffusion along microtubules. This justifies the very good agreement between one-dimensional diffusion models  and the observed experimental data (the motion of mRNA is observed along the embryo wall). Two and three-dimensional diffusion models would predict protein concentrations in the interior of the embryo, which is not observed.

\section*{Materials and Methods}

To fit the sets of data points of Figure~\ref{fig2},
we   consider  that each data set  is approximated by a function $B(x;\vec \alpha)$, with $x\in[0,1]$ and where $\vec \alpha=(\alpha_1, \ldots , \alpha_m)$ is the set of $m$ parameters to be determined. We  assume that the parameter space $S=\{ \vec \alpha : \infty <m_i\le \alpha_i\le M_i<\infty, i=1,\ldots m \}$ is a compact subset of $R^m$. For each fixed value of the vector parameter $\vec \alpha $, and experimental data points $\{(x_i,B_{exp}(x_i))\}_{i=1}^n$, we   consider  the fitness function (sum of the mean squared deviations), 
\begin{equation}
\chi^2(\vec \alpha)= \frac{1}{n}\sum_{i=1}^{n}(B(x_i;\vec \alpha)-B_{exp}(x_i))^2
\label{eq3.1}
\end{equation}
The set of parameter values  that best fits the experimental data is determined from the global minimization condition,
$$
\min_{\vec \alpha \in S}\chi^2(\vec \alpha)
$$

In order to search the global minimum of the function $\chi^2(\vec \alpha)$, with   $\vec \alpha \in S$, we   take  a set of $p$ vectors $\vec\alpha_k$, with $k=1,\ldots ,p$, randomly equidistributed in
the set $S$. Then, for each $\vec\alpha_k$, we compute the fitness function (\ref{eq3.1}). 

To search for the global minimum of the mean squared deviation $\chi^2(\vec \alpha)$, we swarm the set of $p$ vectors $\vec\alpha_k$ in the parameter space $S$. For each vector $\vec\alpha=(\alpha_1,\ldots ,\alpha_m)$, we construct a new vector $\vec\alpha'$ according to the swarm rule,
\begin{equation}
\alpha_i'=\alpha_i+\Delta t (M_i-m_i)(2\xi-1)
\label{eq3.3}
\end{equation}
where $\Delta t$ is a time parameter, $(M_i-m_i)$ is a scaling constant, and $\xi$ is a random variable uniformly distributed in the interval $[0,1]$. Then, we recalculate the new value of the fitness function, $\chi^2(\vec \alpha')$. If $\chi^2(\vec \alpha')<\chi^2(\vec \alpha)$, the  parameter value $\vec \alpha$ is updated to the new value $\vec \alpha'$. If $\chi^2(\vec \alpha')\ge \chi^2(\vec \alpha)$, no update is done. We repeat this procedure for all the parameter values in the search space $S$.

After iterating the swarm  algorithm  $M$ times for all the population of parameter values, we order the 
parameter vectors according to their fitness values, and we discard half of the parameters that have the worst fitnesses. We repeat this procedure $s$ times. The parameter values that best fit the experimental data are the ones that corresponds to the minimum of $\chi^2(\vec \alpha)$. 

This simple algorithm relies on the assumption that the initial number of random points are equidistributed in $S$ and they form a sufficiently dense set in $S$.

In the cases in Figure~\ref{fig2}, the convergence of the swarm algorithm for the determination of the global minimum of the fitness function (\ref{eq3.1}) as a function of the parameters has been checked by graphical methods in two-dimensional sections of the parameter space $S$. 
In all the fits in Figure~\ref{fig2}, the best convergence has been obtained with the swarm parameters, $p=1024$, $M=500$, $s=5$ and $\Delta t=0.01$, and good convergence for the global minima has been obtained.

A further extension of this calibration technique using evolutionary algorithms for the Bicoid-Caudal protein regulation in \textit{Drosophila} has been developed in \citep{Dilao}.

\section*{Acknowledgments}
We thank Fred Cummings, Ana Pombo and Solveig Thorsteinsdottir for enlightening discussions during the preparation of this paper. This work has been supported  by European project GENNETEC, FP6 STREP IST 034952.

\end{document}